\documentclass{PoS}

\usepackage{epstopdf}
\usepackage[utf8x]{inputenc}
\usepackage[T1]{fontenc}
\usepackage{amsmath}
\usepackage{slashed}
\usepackage{graphicx}
\usepackage{subcaption}
\usepackage{simplewick}
\usepackage{verbatim}

\graphicspath{ {../} }
\title{Phase structure of $\mathcal{N}$=1 super Yang-Mills theory from the gradient flow }

\ShortTitle{Gluino condensate from the gradient flow}

\author{Georg Bergner, $^{a}$ \speaker{Camilo Lopez}$^{a}$, and Stefano Piemonte$^{b}$ \\
\llap{$^{a}$} University of Jena, Institute for Theoretical Physics\\ Max-Wien-Platz 1, D-07743 Jena, Germany \\
\llap{$^{b}$} University of Regensburg, Institute for Theoretical Physics \\ Universit\"atstr. 31, D-93040 Regensburg, Germany\\
        E-mail: \email{georg.bergner@uni-jena.de}, \email{camilo.lopez@uni-jena.de}, \email{stefano.piemonte@ur.de}}

\abstract{Composite operators of bare fermion fields evolved along a trajectory on field space by means of flow equations are multiplicatively renormalized. Therefore, even in the case of Wilson fermions, the renormalization of expectation values of fermion operators can be drastically simplified on the lattice. We measure the gluino condensate in $\mathcal{N}$=1 supersymmetric Yang-Mills theory at non-zero temperatures by means of the gradient flow. The non-vanishing expectation value of the gluino condensate up to a certain critical temperature is a signal of chiral symmetry breaking, in agreement with theoretical conjectures on the vacuum structure of the theory. Furthermore, the deconfinement phase transition seems to occur close to this critical temperature, meaning that in $\mathcal{N}$=1 SYM the phases of broken chiral symmetry and of confinement would coincide.}

\FullConference{The 36th Annual International Symposium on Lattice Field Theory - LATTICE2018\\
		22-28 July, 2018\\
		Michigan State University, East Lansing, Michigan, USA.}

\begin{document}

\section{Introduction}

Confinement and chiral symmetry breaking are two relevant non-perturbative phenomena occurring in many gauge theories in four dimensions. Confinement is the emergent property of strong interactions, responsible for the binding of gluons and fermions in complex colorless composite objects like glueballs, mesons, and baryons. Asymptotic freedom implies that the fundamental degrees of freedom of gauge theories appear as single particles only at very high energy, but they are an inadequate description of the physical spectrum at low energy. Confinement is related to the spontaneous breaking of the center symmetry of the gauge group. Chiral symmetry breaking is related to the low-energy dynamics of Yang-Mills (YM) theories interacting with fermions. The intact symmetry of the quantum theory under chiral rotations of the fermion fields is in many cases spontaneously broken by a non-vanishing expectation value of the fermion condensate. Both phenomena, chiral symmetry breaking and confinement, are important features determining the particle spectrum of YM theories.

The relation between the low and high energy regimes of YM theories can be investigated as a function of the temperature. A phase transition is expected each time a symmetry of the theory is broken or restored. A crucial question arising from the two phenomena described above is if the critical temperatures of confinement and chiral symmetry breaking coincide, and if both share the same non-perturbative origin. This question is far from trivial. Although QCD, for example, is characterized by confinement and broken chiral symmetry at low energies, some theories have been found to exhibit only one of those \cite{Aharony:2006da}. Numerical studies of QCD have been able to show that there is approximately a narrow range of temperatures where chiral symmetry restoration and deconfinement take place. However, in QCD these phase transitions are actually crossovers due to the fact that quarks, being color charges transforming in the fundamental representation of the gauge group, break center symmetry and have a non-zero mass which breaks chiral symmetry explicitly even at the classical level. Therefore there are no exact order parameters in QCD which unambiguously define the critical temperature.

The $\mathcal{N}=1$ supersymmetric Yang-Mills theory (SYM) is different from this perspective, both center and chiral symmetry are defined by exact order parameters. The gluinos are interacting with the gluons in the adjoint representation of the group and center symmetry is exact. Gluinos, being the fermionic superpartners of the gluons, are massless and chiral symmetry is not explicitly broken. We can therefore study the $\mathcal{N}=1$ SYM to investigate the non-perturbative origin of confinement and chiral symmetry breaking in a theory where both phenomena are manifest. A strong relation between the two is expected if at least both critical temperatures are coincident, especially if they are driven by the same mechanisms.

Lattice numerical simulations are an import tool for the non-perturbative investigation of the phase diagram of a theory. There are however many issues concerning renormalization and chiral symmetry on the lattice, arising from the Nielsen-Ninomiya theorem. The explicit breaking of chiral symmetry for Wilson fermions requires both additive and multiplicative renormalization
\begin{align*}
\langle\lambda\lambda\rangle_{\mathrm{R}}=Z_{\lambda\lambda}(\beta)(\langle \lambda\lambda\rangle_{\mathrm{B}}-\mathbf{b_{0}})\,.
\end{align*}
In previous investigations \cite{Bergner:2014saa} the additive constant was removed by subtracting the bare condensate at zero temperature,
\begin{align*}
\langle\lambda\lambda\rangle_{\mathrm{S}}=\langle\lambda\lambda\rangle_{\mathrm{B}}^{T=0}-\langle\lambda\lambda\rangle_{\mathrm{B}}^{T}\; .
\end{align*}
The drawback of this approach is that the subtracted condensate is fixed to vanish at $T=0$. Although this subtraction should preserve the behavior of the order parameter near the critical temperature, it is not possible to determine the renormalized condensate at zero temperature and the determination of $\langle\lambda\lambda\rangle_{\mathrm{B}}^{T=0}$ leads to additional uncertainties. No statement can be made regarding the conjecture of having a non-vanishing chiral symmetry breaking condensate at zero temperature. In this work it will be discussed how the additive renormalization is removed, if the fields are flowed using a certain heat-kernel equation. This procedure allows to measure the gaugino condensate even at vanishing temperature. In section \ref{sec:phasediagram} the realization of confinement and chiral symmetry in SYM is briefly reviewed. Then, in section \ref{sec:gradientflow} the gradient flow is shortly discussed as well as the flowed gaugino condensate. Finally, numerical results are presented in section \ref{sec:results}.

\section{Phase diagram of $\mathcal{N}=1$ SYM}\label{sec:phasediagram}

\subsection{Confinement}

Confinement is formally understood in terms of the Polyakov loop, the path-ordered product of the gauge-links wrapping around in the time direction, representing the potential of a single static isolated color charge. As such, the Polyakov loop is not invariant under center symmetry transformations where the time-like gauge links in given time-slice are multiplied by $\exp\left\{(2\pi n \textrm{i})/N_c\right\}$, being $n\in\{0,1,\dots N_c - 1\}$. A vanishing vacuum expectation value of the Polyakov loop signals an unbroken centre symmetry and an infinitely large free energy for isolated static fundamental quarks. At some critical temperature, centre symmetry is broken and the energy of an isolated quark is finite: the theory enters in the deconfined phase, going from a phase of magnetic disorder to one of magnetic order. Confinement is a phase of magnetic disorder where centre symmetry is unbroken. Thus, the gauge theory near the deconfinement phase transition is similar to the Ising model. By adding a massless adjoint Majorana fermion YM becomes minimally supersymmetric. Would be this theory still confining? Indeed, the presence of adjoint matter does not represent a complication: because of its N-ality, it transforms in the same way under centre transformations as the gauge fields. The theory remains thus invariant with respect to centre transformations and the previous analysis is still valid. A phase transition is then expected at some finite temperature and the Polyakov loop still represents a good order parameter \footnote{The behavior of the order parameter near the critical point can be understood through, for example, the Svetitsky-Yaffe conjecture \cite{Svetitsky:1982gs}}. 

\subsection{Chiral symmetry}

Classically, chiral symmetry coincides with the U(1) R-symmetry of the theory, which is part of the SUSY algebra 
$\lambda\to\lambda' = \exp(-i\omega\gamma_{5})\lambda$. At the quantum level this symmetry is anomalously broken by instanton contributions, as it can be proven by e.g. computing the triangle Feynman diagram with one U(1) current and two gauge fields. Taking into account the $\theta$-term in the SU($N$) SYM action, the U(1) chiral rotation is equivalent to a shift $\theta_{\mathrm{YM}}\to \theta_{\mathrm{YM}}-2N\alpha$, where $N$ is the number of colors. The path integral is invariant only for $\alpha=k\pi/N$, i.e.\ the U(1) chiral symmetry is broken down to the subgroup Z$_{2N}$. An interesting question is if, as in the case of QCD, the subgroup which is not broken by the anomaly is spontaneously broken by some fermion condensate. The non-vanishing expectation value of the gluino condensate \cite{Shifman:1987ia,Morozov:1987hy} \footnote{From holomorphy of the superpotential, anomaly matching and symmetry it is known that the bi-fermion condensate $\langle \lambda\lambda\rangle$ is non-zero \cite{Terning:2006bq, Schwetz:1997cz}.} ($\langle \lambda\lambda\rangle$ ) is not invariant with respect to the full Z$_{2N}$ group since $\langle \lambda \lambda \rangle = \mathrm{e}^{2i\alpha}\langle \lambda\lambda\rangle$ only if $\alpha=0,\pi$. Thus, the global Z$_{2N}$ symmetry is spontaneously broken down to the subgroup Z$_{2}$ and the discrete chiral symmetry is then broken at zero temperature, where $N$ degenerated vacua coexist. This coexistence suggests that the transition to the broken phase as a function of the gluino mass is first order. Furthermore, because the symmetry breaking pattern is Z$_{2N}\to$Z$_{2}$, the theory is at the quantum level very similar to the Z$_{N}$ Ising model. This gives rise to the idea that the first order phase transition should terminate at a second-order end point at some critical temperature $T_{\mathrm{crit}}$, where the Z$_{2N}$ chiral symmetry should be recovered.

\section{The gradient flow and the condensate}\label{sec:gradientflow}

Motivated in the context of trivialising maps \cite{Luscher:2009eq}, L\"uscher studied the correlation functions of fields \textit{flowed} through the equations \cite{Luscher:2011bx} \footnote{The term $D_{\mu}\partial_{\nu}B_{\nu}$ is a gauge parameter, which is included for mere technical reasons.}
\begin{align}\label{gaugeflow}
&  \partial_{t}B_{\mu}=D_{\nu}G_{\nu\mu} + D_{\mu}\partial_{\nu}B_{\nu},\quad \left.B_{\mu}\right|_{t=0}=A_{\mu},\quad
G_{\mu\nu}=\partial_{\mu}B_{\nu}-\partial_{\nu}B_{\mu}+\left[ B_{\mu},B_{\nu} \right]\\
& \partial_{t}\chi=\Delta \chi, \quad \partial_{t}\bar{\chi}=\bar{\chi}\overleftarrow{\Delta},\quad \left.\chi\right|_{t=0}=\psi, \quad \left.\bar{\chi}\right|_{t=0}=\bar{\psi},\quad  \Delta=D_{\mu}D^{\mu}.\nonumber
\end{align}

where $B$ and $G$ are the Yang-Mills gauge field and field strength, respectively. Moreover $\chi$ is a spinor field and $D_{\mu}$ is the gauge-covariant derivative in the adjoint representation. The parameter $t$ describes a flow on the space of gauge fields. These equations have a smoothening effect on the fields, which are Gaussian-like smeared over an effective radius $r_{t}=\sqrt{8t}$, as it can be easily seen by integrating equation (\ref{gaugeflow}) in the non-interacting limit $[B,B]\sim 0$
\begin{align*}
B_{\mu,1}(t,x)=\int{d^{D}y~K_{t}(x-y)A_{\mu}(y)}, \quad K_{t}(z)=\int{\frac{d^{D}p}{(2\pi)^{D}}\mathrm{e}^{ipz}\mathrm{e}^{-tp^{2}}}=\frac{\mathrm{e}^{-z^{2}/4t}}{(4\pi t)^{D/2}},
\end{align*}

where $D$ is the number of space-time dimensions and $K$ the heat-kernel. Moreover, the term $\mathrm{e}^{-tp^{2}}$ regularizes the integral in momentum space when $t>0$, removing the UV divergences at large momenta. Some years ago it was shown in \cite{Luscher:2011bx} that through this kind of gradient flow the smearing property remains at all orders. It was also found by BRS-symmetry, that correlation functions of monomials of flowed gauge-fields are renormalized without the necessity of extra counter-terms and spinor operators renormalize multiplicatively \footnote{The monomials renormalize according to the field content since the flowed fields are effectively non-local.}  \cite{Luscher:2013cpa}. Moreover, one specific advantage of the gradient flow relies on the fact that it is regularization-scheme independent. Therefore, all these results are expected to hold also on a space-time lattice. Thus, currents and densities which are explicitly broken by the lattice realization should be more accessible within this method. As a special case, the additive renormalization constant, necessary for the computation of the gaugino condensate with Wilson fermions, is not required if these are flowed up to some finite flow-time. Consequently, it is possible to study if $\langle \lambda\lambda \rangle\neq 0$ at $T=0$ even with Wilson fermions. The value of Z$_{\lambda\lambda}$
is irrelevant if the bare lattice gauge coupling is fixed. Non-zero temperatures are achieved by compactifying and changing the number of lattice sites in temporal direction, where fermion fields fullfil anti-periodic boundary conditions. The flowed bare gaugino condensate can be defined as \cite{Luscher:2013cpa}
\begin{align}\label{contcond}
\langle \bar\lambda(x,t)\lambda(x,t)\rangle_{\mathrm{B}}=-\int{d^{D}v d^{D}w~K(t,x;0,v)S(v,w)K(t,x;0,w)^{\dagger}}.
\end{align}
Hence, computing the flowed condensate amounts to the action of the heat-kernel on the fermion propagator $S$. The chiral condensate can be calculated on the lattice by means of the discrete version of equation (\ref{contcond})
\begin{align*}
\langle \bar\lambda(x,t)\lambda(x,t)\rangle_{\mathrm{B}}=-\sum_{v,w}\langle{\mathrm{tr}\{K(t,x;0,v)(D_{\mathrm{W}}(v,w))^{-1}K(t,x;0,w)^{\dagger}\}\rangle},
\end{align*}

where $D_{\mathrm{W}}$ is the Wilson-Dirac operator and the trace runs over spinor and gauge group indices. Following \cite{Luscher:2013cpa}, the trace is estimated stochastically by inserting a complete set of random complex spinor fields $\eta(x)$ with $\langle \eta(x) \rangle=0$ and $\langle \eta(x)\eta(y)^{\dagger}\rangle=\delta_{x,y}$:

\begin{align*}
\langle\bar\lambda\lambda(t) \rangle=\frac{1}{N_{\Gamma}}\sum_{x\in \Gamma}{\langle \bar\lambda\lambda(x,t)}\rangle=-\frac{1}{N_{\Gamma}}\sum_{v,w}\langle \xi(t;0,v)^{\dagger}D^{-1}_{\mathrm{W}}\xi(t;0,w)\rangle, \quad \xi(t;s,w)=\sum_{x}K(t,x;s,w)^{\dagger}\eta(x).
\end{align*}

Finally, to compute the new vectors $\xi$, the so-called \textit{adjoint} flow equation 

\begin{align}
\label{adjfloweq}
\partial_{s}\xi(t;s,w)=-\Delta(V_{s})\xi(t;s,w), \quad \xi(t;t,w)=\eta(w),
\end{align}

must be integrated from $s=t$ to $s=0$, i.e. backwards in comparison with the flow equations presented above. Here, the gauge connection $V_{s}$ in the covariant Laplacian $\Delta$ is flowed up to the same $t$ as $\eta$.

\section{Results and outlook}\label{sec:results}

\subsection{SU(2)}

The first investigations focused on the phase diagram of SYM with SU(2) gauge group and are built upon the results of \cite{Bergner:2014saa}. For this purpose, three different lattice ensembles were chosen with coupling $\beta=1.75$ and Hopping parameter $\kappa\in \{0.1480,0.1490,0.14925\}$. The lattice size was set to $24^{3}\times N_{t}$ for $N_{t}\in \{4,5,6,\cdots,48\}$, with the upper bound parametrizing the zero-temperature limit. Setting the scale, i.e. going from lattice to physical units, was achieved through the $t_{0}$ parameter, which is defined through the gradient flow as \cite{Luscher:2010iy}\footnote{Here a mass-dependent scheme was chosen, i.e. different t$_{0}$ were taken for different $\kappa$ value.}

\begin{align*}
  \left.t^{2}\langle \varepsilon(t)\rangle\right|_{t=t_{0}}=0.3, \quad \varepsilon=\frac{1}{4}G_{\mu\nu}^{a}G_{\mu\nu}^{a},
\end{align*}

with $\varepsilon$ the field energy density and $G_{\mu\nu}^{a}$ the flowed curvature tensor.

\begin{figure}[h!]
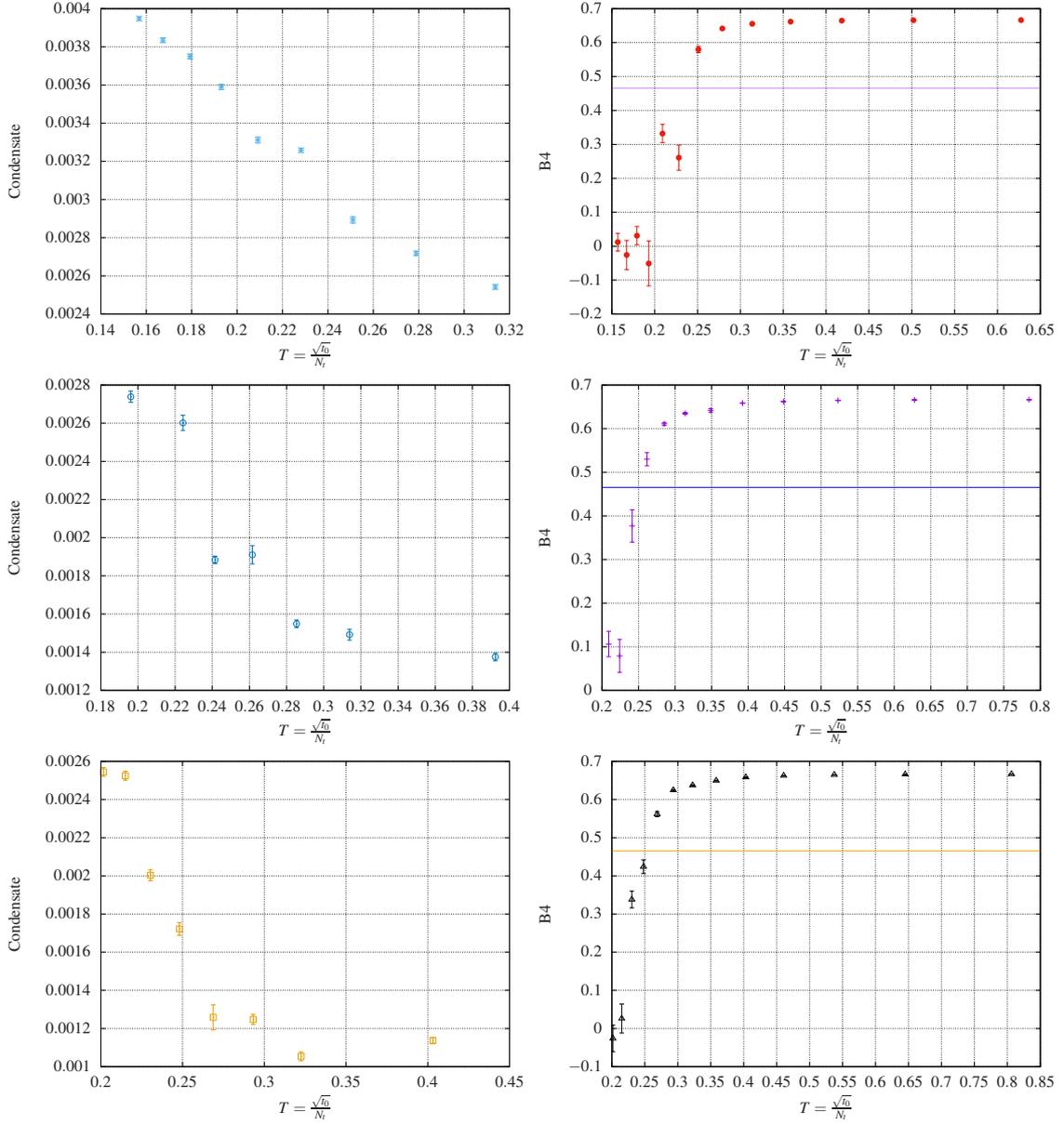

  \centering
  \captionsetup{font=footnotesize} 
  \begin{subfigure}[b]{0.49\linewidth}
    \scalebox{.6}{\input{./condensate-t0.tex}}
  \end{subfigure}
  \begin{subfigure}[b]{0.49\linewidth}
    \scalebox{.6}{\input{./B41480.tex}}
  \end{subfigure}

  \begin{subfigure}[b]{0.49\linewidth}
    \scalebox{.6}{\input{./condensate1490.tex}}
  \end{subfigure}
  \begin{subfigure}[b]{0.49\linewidth}
    \scalebox{.6}{\input{./B41490.tex}}
  \end{subfigure}

  \begin{subfigure}[b]{0.49\linewidth}
    \scalebox{.6}{\input{./condensate14925.tex}}
  \end{subfigure}
  \begin{subfigure}[b]{0.49\linewidth}
    \scalebox{.6}{\input{./B414925.tex}}
  \end{subfigure}
  \caption{On the left side the temperature dependence of the bare gluino condensate is shown for $\kappa=0.1480,1490$ and $0.14925$, with the smallest $\kappa$ in the uppermost plot. On the undermost graphic, the point corresponding to the highest temperature appears to show a growth in the condensate. This is however a non-physical oversmoothing artifact due to the fact that $\sqrt{8t_{0}}> N_{t}$ in that region. The right-hand side shows the Binder cumulant of the Polyakov loop for the same $\kappa$ values.   }
  \label{su2}
\end{figure}

The results for SU(2) are summarized in Fig.~\ref{su2}. It can be seen that for lower $\kappa$ values, i.e. for large gaugino mass, the condensate tends to lower values as the temperature increases, although a phase transition is difficult to identify. This is of course expected, since the "jump" of the order parameter around the pseudo-critical temperature is much more smoothen out when the mass, i.e. the parameter which explicitly breaks the symmetry, is far away from zero. For the smallest gaugino mass, i.e. $\kappa=0.14925$ the signal is considerably better and the jump at the critical temperature is more pronounced. It is remarkable that the phase transition appears to happen at $T\sim 0.25$ for all $\kappa$ values. This temperature approximatly coincides with the deconfinement phase transition, which occurs when the Binder cumulant of the Polyakov loop reaches the critical value shown on the right side of Fig.~\ref{su2}. In Ref.~\cite{Bergner:2014saa}, it was shown that the deconfinement phase transition is second order and has the critical behavior of the $Z_{2}$ Ising model. The signal for the phase transition persists in the chiral continuum limit. This is an indication that in SU(2)  $\mathcal{N}=1$ supersymmetric Yang-Mills theory the deconfinement phase transition and restoration of the discrete $Z_{2N_{c}}$ chiral symmetry occur simultaneously. In this aspect, SYM theory seems to be more similar to the conjectures about QCD. Other investigations of adjoint QCD-like theories have found, on the contrary, mixed deconfined-non-chiral phases~\cite{Karsch:1998qj,Bilgici:2009jy,Braun:2012zq}. Because of the uncertainties, these results should however be taken carefully, more precise measurements are required and are ongoing in the near future. The results could be improved by simultions near $m_{g,\mathrm{R}}=0$ on large lattice volumes, taking more configurations near the critical temperature.

\subsection{Preliminary results for SU(3)}

\begin{figure}[h]
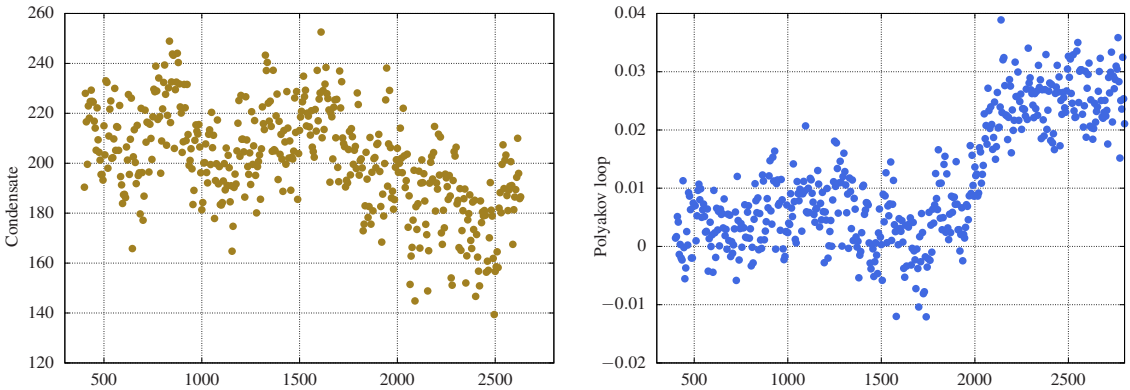

  \centering
  \captionsetup{font=footnotesize}
  \begin{subfigure}[h]{0.49\linewidth}
    \scalebox{.6}{\input{./condhis.tex}}
  \end{subfigure}
  \begin{subfigure}[h]{0.49\linewidth}
    \scalebox{.6}{\input{./plhis.tex}}
  \end{subfigure}
  \caption{Monte-Carlo histories of the chiral condensate (left) and the Polyakov-loop (right).}
  \label{su3}
\end{figure}

Currently, finite-temperature studies are running in order to study the phase diagram of SU(3) SYM. Up to now both, the Polyakov loop and the chiral condensate, have been measured without gradient flow on a 16$^{3}\times 9$ lattice with $\beta=5.50$, $\kappa=0.1673$. This $N_{t}$ corresponds to the estimates critical deconfinement temperature. In Fig.~\ref{su3} the Monte-Carlo histories of the gluino condensate and the Polyakov-loop are shown. It can be seen that as the former decreases, the latter acquires a non-vanishing value. Indeed the Pearson coefficient is found to be $\rho=-0.565$, what can be interpreted as a negative correlation between both order parameters. Although this is just a preliminary result, it is already remarkable and gives the hint that coincident critical temperatures are also to be found in the case of SU(3). Currently further simulations are running and we are also planning to include the gradient flow analysis of the condensate as in the case of SU(2).

\section*{Acknowledgments}
We thank the members of the DESY-Münster collaboration Sajid Ali, Istvan Montvay, Gernot M\"unster and Philipp Scior for helpful discussions, comments, and advises.
The authors gratefully \linebreak acknowledge the Gauss Centre for Supercomputing
e.~V.\,(www.gauss-centre.eu) for funding this project by providing
computing time on the GCS Supercomputer JUQUEEN and JURECA at J\"ulich
Supercomputing Centre (JSC) and SuperMUC at Leibniz Supercomputing Centre
(LRZ). G.~B.\ and C.~L.\ acknowledge support from the Deutsche Forschungsgemeinschaft (DFG) Grant
No.~BE 5942/2-1.

\end{document}